# Excitonic Effects in the Optical Spectra of Lithium metasilicate (Li$_2$SiO$_3$)

Nguyen Thi Han[1,*], Vo Khuong Dien[1], Ming-Fa Lin[1,2,*].

*Department of Physics, National Cheng Kung University, 701 Tainan, Taiwan[1]*

*Hierarchical Green-Energy Materials (Hi-GEM) Research Center, National Cheng Kung University, Tainan, Taiwan[2]*

Email: han.nguyen.dhsptn@gmail.com, mflin@mail.ncku.edu.tw.

## Abstract

The Li$_2$SiO$_3$ compound, a ternary electrolyte compound of Lithium-ion based batteries, exhibits unique geometric and band structures, an atom-dominated energy spectrum, charge densities distributions, atom and orbital-projected density of states, and strong optical responses. The state-of-the-art analysis, based on an ab-initio simulation, have successfully confirmed the concise physical/chemical picture and the orbital bonding in Li-O and Si-O bonds. Additionally, the unusual optical response behavior includes a large red shift of the onset frequency due to the extremely strong excitonic effect, the polarizations of optical properties along three-directions, 22 optical excitations structures and the most prominent plasmon mode in terms of the dielectric functions, energy loss functions, absorption coefficients and reflectance spectra. The close connections of electronic and optical properties can identify a specific orbital hybridization for each distinct excitation channel. The developed theoretical framework will be very appropriate for fully comprehending the diverse phenomena of cathode/electrolyte/anode materials in ion-based batteries.

# 1. Introduction

Lithium-ion based batteries (LIBs) are one of the main-stream materials in basic science and play important roles in life and technology since they have many outstanding properties namely low cost, lightweight, long lifetime, rapid charging/discharging, and work with a voltage of the order of 4V with specific energy ranging between 100 Whkg$^{-1}$ and 150 Whkg$^{-1}$ [1-5]. However, recent commercial LIBs with liquid electrolytes contain potential security risks concerning volatilization, flammability, and explosion. Currently, experimental and theoretical studies evidence that secondary batteries using solid electrolytes would overcome these safety issues. Additionally, the battery cell design with the solid states would be more simplified, providing a wider electrochemical window than liquid ones [6-10]

Some potential candidates for solid-state electrolyte materials can be listed, such as $Li_3ClO$, $Li_2SiO_3$, $Li_2PS_3$ [11-13], and so on. Among them, the $Li_2SiO_3$ compound has gained considerable attention due to its special performance in various fields. For example, the ternary $Li_2SiO_3$ compound with low-cost synthesis exhibits a reliable ionic conductivity ($2.5 \times 10^{-8}$ S.cm$^{-1}$ at 25 °C) [14] and high mechanic strength, which is suitable for a fast and flexible thin-film ionic conductor [15], and a $CO_2$ fine absorption both effect evidence that this material can be utilized for gas captures [16]. The polar orthorhombic symmetry of the orthorhombic structure suggests that this material can be suitable for piezoelectric, pyroelectric, and electro-optic applications. Furthermore, recent studies also indicate that the $Li_2SiO_3$ compound also has been actively investigated as a luminescent material [17,18] due to its outstanding optical properties.

Up till now, Lithium metasilicate has been successfully synthesized by various methods [17-21]. Its essential physical/chemical properties, such as optimized geometry structures and surface monologies, are usually examined by X-ray diffraction [17,18,22], scanning electronic microscopy (SEM) [23] and tunneling electronic microscope (TEM) [24,25]. The energy spectrum depending on frequencies and wave vectors at the valence states and the van Hove singularities in the density of states are not achieved because of a lack of angle-resolved photoemission spectroscopy (ARPES) [26] and scanning tunneling spectroscopy (STS) measurements [27], respectively. Regarding the optical responses, photoluminescence, absorption, and transmission measurement have been done conducted [17,18,28]. However, they have only been investigated for doped $Li_2SiO_3$ or measured at high temperature, in which optical properties are strongly affected by surfaces defects.

On the other hand, in spite of the previous theoretical investigations that have investigated the electronic properties, transport mechanisms and the optical properties of the $Li_2SiO_3$ compound [29-31], most studies do not propose a concise physical/chemical picture that closely relates to the performance of the energy storage system. Our previous studies have linked the orbital hybridizations with the geometric, electronic properties of this material [32]. However, the systematic combination of the geometric, electronic, and excitonic effects on the optical response with the energy-dependent orbital hybridizations has not been achieved so far.

In this Letter, the theoretical framework, which is established on the significant orbital hybridizations, is utilized to examine the crucial features in electrolyte materials of $Li^+$ related batteries. This strategy is based on the density functional theory calculation on the optimized geometric structure with position-dependent chemical bonding, the atom dominated energy spectrum at various energy ranges, the spatial charge distribution due to different orbitals, and the atom- and orbital-projected density of states associated with the overlap of orbitals. The specific orbital hybridizations will be used to interpret the onset of the optical frequency, stable excitonic states, a lot of prominent absorption structures, a very strong plasmon mode in terms of the dielectric functions, energy loss functions, absorption coefficients, and reflectance spectra under the distinct electric polarization. The predictions in this work require highly-resolved experimental measurements.

## 2. Computational details

We have used the density functional theory (DFT) method via the Vienna Ab-initio Simulation Package (VASP) [33] to perform the optimization of the structure and calculated the electronic and optical properties. The Perdew-Burke-Ernzerhof (PBE) generalized gradient approximation was used for the exchange-correlation functional [34]. The projector augmented wave (PAW) method was utilized for characterization of the ion and valence electron interactions [35]. The cutoff energy for the expansion of the plane wave basis was set to 550 eV. The Brillouin zone was integrated with a special k-point mesh of 8×8×8 in the Γ-centered sampling technique [36] for the structural optimization. The convergence condition of the ground-state is set to be $10^{-8}$ eV between two consecutive simulation steps, and all atoms were allowed to fully relax during the geometric optimization until the Hellmann-Feynman force acting on each atom was smaller than 0.01 eV.

In this paper, the optical properties of the $Li_2SiO_3$ compound have been investigated by adopting the method of Rohlfing and Louie [37]. Following the DFT results, the quasiparticle energies are obtained within GW approximation for the self-energy [38], the cutoff energy for the response function was set to 220 eV, 8×8×8 Γ-centered kpoints sampling was used to represent reciprocal space. The wannier interpolation procedure performed in the WANNIER90 code [39] was used to plot the quasi-band structure. Regarding the optical properties involve in the excitonic effect, the standard Bethe-Salpeter equation (BSE) in the Tamm Dancoff approximation was employed to take the electron-hole interactions into account [40]. The k-point sampling, energy cutoff, number of bands, were set to the same values as in the GW calculation. In addition, the 8 lowest unoccupied conduction bands (CBs) and the 20 highest occupied valence bands (VBs) are included as a basis for the excitonic states with a photon energy region from 0 eV to 28 eV. In addition, the Lorentzian broadening γ, which arises from various de-excitation mechanisms, was set at 0.1 eV for all optical calculations. Other optical properties, such as the energy loss function, absorption coefficient, and reflectivity can be obtained from the dielectric function by the following relations:

$$L(\omega) = \frac{\varepsilon_2(\omega)}{\varepsilon_1^2(\omega)+\varepsilon_2^2(\omega)}, \quad R(\omega) = \left|\frac{\sqrt{\varepsilon_1(\omega)+\varepsilon_2(\omega)}-1}{\sqrt{\varepsilon_1(\omega)+\varepsilon_2(\omega)}+1}\right|^2, \quad \alpha(\omega) = \sqrt{2}\omega\left[\sqrt{\varepsilon_1^2(\omega)+\varepsilon_2^2(\omega)}-\varepsilon_1(\omega)\right]^{1/2}$$

## 3. Results and discussions.

In this Letter, we have chosen to investigate the electronic and optical properties of the $Li_2SiO_3$ compound. This solid-state electrolyte crystallizes in an orthorhombic structure with space group [Cmc21] **(Fig. 1(a))**. The calculated lattice constants of this compound are 9.467 Å, 5.440 Å and 4.719 Å for x-, y- and z-directions, respectively, which are close to the value in recent experimental [41] and theoretical investigations [42]. The conventional $Li_2SiO_3$ cell contains 24 atoms (8-Li, 4- Si and 12-O atoms), where each Lithium/Silicon atom is tetrahedrally coordinated and bonded to four oxygen atoms, **(Fig. 1(b))**. The basic structural unit is comprised of 32 Li-O and 16 Si-O chemical bonds, in which the bond length fluctuation ranges are ~ 1.94–2.20 Å and ~ 1.61–1.70 Å for the former and the latter, respectively. Obviously, the extremely non-uniform chemical/physical environment indicates is very useful for supporting the $Li^+$ migration in electrolyte materials [43]. Furthermore, this behavior also is responsible for the anisotropic optical properties. The unusual phenomena presented here evidence the complex nature of the orbital-hybridizations in the chemical bonding.

The unusual physical/chemical environment is generated by the complex orbital hybridizations and directly reflected in the electronic structure. To examine this property, the energy spectrum along high symmetry points **(Fig.1(c))** of this material is calculated and shown in **Fig. 2 (a).** As can be seen from the diagram, the $Li_2SiO_3$ compound displays rich and unique electronic properties, a high asymmetry of the valence hole and the conduction electron states were observed within a wide range of energy (-28 eV – 13 eV). The presence of many energy sub-bands with complicated energy dispersion, such as parabolic, linear, oscillatory or almost flat energy bands, are a result of the contribution of many atoms and orbitals in the large Moire cell. Additionally, band edge-states which are the extreme or saddle points in the energy-wave-vector, are situated at the highly symmetric points and the other wave vector between them. These critical points in the energy spectrum will create strong van Hove singularities and therefore are responsible for the prominent peaks in the optical absorption spectrum. Very interestingly, the highest occupied state and the lowest unoccupied state, respectively, are located at the Z and Γ points, leading to a wide indirect insulator gap of 8.4 eV. This indirect band gap is located very close to the HSE06 calculated band gap [44]. The mentioned results are associated with the complex orbital hybridizations in the chemical bonds, the close combinations of which with the charge density distribution, the atom and orbital projected-density of states, the specific orbital hybridizations in chemical bonds, are related to the prominent absorption structures, could be successfully achieved.

In addition to the band structures, the atom dominated-energy spectrum also provides partly useful information about orbital hybridizations. For the entire band structure, the Li atoms' contribution is negligible. Yet they cannot be ignored since in the absence of the Li-O chemical bonds, some noticeable features will disappear. According to the domination of the various atoms and orbital as shown in **Figs. 2(b), 2(c) and 2(d),** the electronic band structure could be systematically divided into five sub-groups: (I) $E^c > 8.5$ eV, (II) $4.2$ eV $< E^c < 8.5$ eV, (III) $-10.5$ eV $< E^v < -4$ eV, (IV) $-13$ eV $< E^v < -10.5$ eV, and (V) $-28$ eV $< E^v < -23$ eV. Obviously, the V group below $-22$ eV is separated by a huge gap from the IV group. Based on the obvious features, each energy sub-group could be revealed as the dominance of (I) (Si, O) atoms, (II) (Si, O) atoms (III) (O) atom, (IV) (Si, O) atoms and (V) (Si, O) atoms. Through the combination of these dominant parts with the charge density distribution, the atom and orbital-projected density of states, the fully the concise chemical/physical pictures in Li-O, Si-O chemical bonds could be identified.

In order to generalize the orbital hybridization between the constituents, the charge density distribution of the Li, Si and O atoms before/after modification were performed. As presented in **Fig. 3(a),** the localizations of the charge on isolated Li atom (the yellow and pink parts), correspond to the 1s and 2s orbitals. The similar but wider charge accommodation is associated with (1s, 2s) and ($2p_x$, $2p_y$, $2p_z$) orbitals of the O atoms **(Fig. 3(c)).** After combination to form the chemical bonds, the inner/outer regions of the Li and O atoms of the Li-O chemical bonds along three electric-polarizations **(Figs 3(d), 3(e) and 3(f))** present the small but finite obvious deformations, especially for the shortest ones. This reflects the presence of non-negligible single (2s-2s) and remarkable multi 2s-($2p_x$, $2p_y$, $2p_z$) orbital interactions between these atoms. As for the Si-O chemical bonds, the highest charge density distribution residing around Si atom arises from the (1s, 2s, $2p_x$, $2p_y$, $2p_z$, 3s) and ($3p_x$, $3p_y$, $3p_z$) orbitals **(Fig. 3 (b))**. Similar but stronger than Li-O, the Si-O bonds are presented by the significant changes of the yellow-pink part combined with the obvious deformation of the pink region **(Figs. 3(g), 3(h), 3(i))** of the Si and O atoms, these propose the significant hybridization of the (3s, $3p_x$, $3p_y$, $3p_z$)-(2s, $2p_x$, $2p_y$, $2p_z$) orbitals.

The orbital character of the Li-O and Si-O chemical bonds, which dominate the essential properties, could be sufficiently elucidated from the atom- and orbital-projected density of states (DOS). The main characteristic of their special structures is governed by the energy dispersions in the energy band structure and the dimensionality of the material. As for the three-dimensional Li$_2$SiO$_3$ compound, the Van Hove singularities mainly come from the extreme or saddle points and the dispersionless of energy sub-bands. This leads to the formation of the shoulder structure and pronounced asymmetric peaks **(Fig 4(a))**. Furthermore, the considerable mixed density of states of difference atoms/orbitals **(Figs. 4(a)-4(d))** clearly evidences the very strong coupling of specific orbital hybridizations. Depending on their co-existence, we can divide five domination parts in turn: (I) Li (2s)-O (2s, 2p$_x$,2p$_y$, 2p$_z$) and Si (3s, 3p$_x$, 3p$_y$, 3p$_z$) – O (2s, 2p$_x$, 2p$_y$, 2p$_z$), (II) Li (2s)- O(2s, 2p$_x$, 2p$_y$, 2p$_z$) and Si (3s, 3p$_x$, 3p$_y$, 3p$_z$) – O (2s, 2p$_x$, 2p$_y$, 2p$_z$), (III) Li (2s) - O (2p$_x$, 2p$_y$, 2p$_z$) and Si (3s, 3p$_x$, 3p$_y$, 3p$_z$) - O (2p$_x$, 2p$_y$, 2p$_z$), (IV) Li (2s) - O (2p$_x$, 2p$_y$, 2p$_z$) and Si (3s, 3p$_x$, 3p$_y$, 3p$_z$)- O (2p$_x$, 2p$_y$, 2p$_z$), and (V) Li (2s) - O (2s) and Si (3s, 3p$_x$, 3p$_y$, 3p$_z$)- O (2s), where each part, respectively, is examined to be associated with each identified sub-group of the energy band structure. This delicate analysis of the orbital characteristics illuminates the excitation mechanism of various strong optical responses.

Following the absorption of a photon with suitable energy, an electron will be vertically excited from an occupied to unoccupied states. The single-particle excitation spectrum can be described by using the linear Kubo formula [45]: $\epsilon_2(\omega) = \frac{8\pi^2 e^2}{\omega^2}\sum_{vc\mathbf{k}}|e\langle v\mathbf{k}|\mathbf{v}|c\mathbf{k}\rangle|^2 \delta(\omega - (E_{c\mathbf{k}} - E_{v\mathbf{k}}))$, where the first part, $|e\langle v\mathbf{k}|\mathbf{v}|c\mathbf{k}\rangle|^2$ is the square of the electric dipole moment, which is responsible for the strength of the excitation peaks, and the second part, $\delta(\omega - (E_{c\mathbf{k}} - E_{v\mathbf{k}}))$, is the joined of the density of states, which corresponds to the available excitation transition channels. In addition to the independent single-particle excitations, the bound state of the electron-hole pair, or so-called exciton is formed through the mutual Coulomb interaction between excited electrons and holes. Under a sufficiently strong attractive interaction, this quasi-particle may be stable enough against collisions with a phonon. The tightly bound exciton is usually observed in large gap materials, strongly affecting the main feature of the optical excitation, and closely related to the critical orbital hybridizations. The physics of the selective optical absorption and the effect of the robust exciton are the main study focus of the current work.

As direct consequence of the electromagnetic (EM) wave absorption, the dielectric function is complex and provides very important information for comprehending the optical transition mechanisms, in which, the prominent responses are due to the electronic excitation and therefore, strongly related to the orbital hybridization. As presented in **Fig 5 (b)**, the single-particle spectrum is described by the imaginary part $\varepsilon_2(\omega)$ of the dielectric function. It presents a serious change under the three different polarizations as a consequence of the non-uniform physical/chemical environment. The threshold frequency, also called the optical gap, around 9.0 eV (**Fig. 5 d**) for independent particle excitations may a little larger than the fundamental indirect gap $E_g^i$ = 8.4 eV due to the conservation of the momentum. However, when the Coulomb interactions are taken into account **(Fig. 5 (b))**, its value is reduced by 2.1/2/2.05 under x-/y-/z-directions. The great redshift indicates very strong electron-hole couplings, and therefore, the excitonic states may be stable at the room temperature.

In addition to the threshold frequency, there also exist an extra 22 non-vanishing shoulders and/or pronounced peaks (the distinct colored arrow). These are associated with the vertical promotion of available valence electrons from the strong van Hove singularities/extreme points in the density of states/band structures and the atom dominated energy spectrum. The presence of an extra two peaks below the band gap is due to the strongly mutual electron-hole interactions at the extreme band edge states. According to the state-of-the-art analysis of the concise physical and chemical pictures in the atom-dominated band structure, atom-/orbital-van Hove singularities, and strong optical responses, the excitation mechanism was achieved and described by the arrowheads in the energy band structure **(Fig.2 a)**. The corresponding chemical characters are denoted by the distinct colors of arrows in the orbital-projected density of states **(Figs. 4 (b), 4(c), 4(d))** and the orbital hybridizations related to these optical responses are successfully identified and shown in **Table 1**. Generally, the majority/minority of peaks of $\varepsilon_2(\omega)$ arise from the electron transition between the occupied O-($2p_x,2p_y,2p_z$) states and the unoccupied Si-($3s, 3p_x, 3p_y, 3p_z$)/Li-2s states. For example, the transition from the O-$2p_z$ to the Si-3s/Li-2s at the Gamma high symmetry point leads to the optical gap at 9 eV in **Figs. 5(d), 5(f), 5(g).** Based on this procedure, the orbital hybridizations associated with the specific optical transitions can be thoroughly understood. This viewpoint could be developed for other anode or cathode compounds of LIBs as well.

The real part $\varepsilon_1(\omega)$ of the dielectric function is related to the imaginary part $\varepsilon_2(\omega)$ through the Kramers-Kronig relationship [46]. Under this relation, the pair pronounced peaks of the real part exist simultaneously with the prominent structures of the imaginary one as present in **Fig. 5(a)**. Within the inactive region, the real part is weakly dependent on the energy. As an example, the background dielectric constants, $\varepsilon_1(0)$, are about 1.9-2.0 for the x-, y- and z- directions and almost stay dispersionless for frequencies less or equal to the optical gap, through which, the low energy reflectance spectrum $R(\omega) = \left|\frac{\sqrt{\varepsilon_1(0)}-1}{\sqrt{\varepsilon_1(0)}+1}\right|^2$ and the vanishing range of absorption coefficients could be identified. Contrarily, during the excitation events, the real part is very sensitive to a change of the photon energy. Very interestingly, at some frequencies, the real part almost vanishes. The simultaneous existence of the zero points of the real part with very weak single-particle transitions indicated corresponds to the weak damping of the plasmon resonances (discussed in detail in **Fig. 6(a)**). For example, the insignificant Landau damping by the single partial excitation at the real part is 19 eV, 18.5 eV and 17 eV for x-, y- and z- directions, respectively. On the other hand, the zero points for the z-direction at 8.1 eV combine with a very strong interband transition peak, and therefore, correspond to the serious Landau damping.

The energy loss function (ELF), being defined as $\text{Im} = \left|\frac{-1}{\varepsilon(\omega)}\right|$ [47] is the screened response function due to the various valence charges of Li, Si and O atoms. Each spectral peak in the ELF corresponds to the excitation of a plasmon mode, a collective excitation of a certain valence charge. As we mentioned before, the zero points in the real part, $\varepsilon_1(\omega)$, is also one way for explaining the coherent behavior of the free electrons. However, due to the combination with the remarkable imaginary $\varepsilon_2(\omega)$ part at certain zero points, the prominent peaks in the ELF may be better in this scenario. As clearly indicated in **Fig 6 (a)**, the strongest peak is located at frequency $\omega_p = $ 19.0/18.5/17.0 eV for the x-/y-/z-direction of electric polarizations due to almost un-damping of the plasmon resonances. These most prominent/important peaks are contributed from the significant orbitals of Li-2s, Si- (3s, 3$p_x$, 3$p_y$, 3$p_z$) and O- (2$p_x$, 2$p_y$, 2$p_z$). Since the contributions mostly appear at very low energy below -22.5 eV and hence, the contribution of the O atom to this mode is not considered. In addition to the most pronounced peak, a few plasmon modes with relatively weak intensity exits owing to the considerable Landau damping.

When a light beam incident on the $Li_2SiO_3$ compound medium along the normal direction, some of the light will be attenuated by the valence electrons and be reflected from the front surface, propagating through a finite-width of the sample. The absorption coefficient and the reflectance of solid-state materials are two general optical phenomena and clearly reflect the principal characteristics of single and coherent excitations. As evident in **Figures 6(b) and 6(c)**, when photon energy is smaller than the onset frequency, due to the lack of the electronic excitation contribution, the absorption coefficient $\alpha(\omega)$ is vanishes, while the reflectance R(ω) is weakly dependent on the frequency. The reflective index is roughly examined by $\left|\frac{\sqrt{\varepsilon_1(0)}-1}{\sqrt{\varepsilon_1(0)}+1}\right|^2$ as we mentioned before. However, beyond the thresh-hold frequency, both $R(\omega)$ and $\alpha(\omega)$ are changing significantly and sensitively to the excitation events. As for the absorption coefficient, the rapid increase is contributed by the various interband transitions, especially, the most obvious change at the plasmon resonance frequency due to the participation of the collective excitations. The inverse values of the absorption coefficient are in the range of 60 Å- 400 Å, indicating that the EM waves propagate in the medium are easily absorbed by the rich electronic excitations. On the other hand, the reflectance is significantly enhanced and displays large fluctuation. The abrupt prominence corresponding to the plasmon mode appears at $\omega_p$. For example, the reflectance has a 22 % variation under the E//xx case (the red curve) at the resonance frequency.

In addition to the geometric structures, the measurement of other features, such as electronic and optical properties require high-resolution techniques. For example, the density of states near the Fermi level can be tested by Scanning tunneling spectroscopy (STS). Unfortunately, this method might only be suitable for a thin-film sample because of the weak quantum current. Angle-resolved photoemission spectroscopy (ARPES) is utilized for examination of the occupied energy band structure. Apparently, the complicated orbital hybridizations in $Li_2SiO_3$ would create a daunting challenge in the examination of the energy sub-bands. Very interestingly, the rich optical properties of the $Li_2SiO_3$ compound, such as the onset of excitonic peaks, a lot of prominent absorption structures, and the strongest coherent excitation mode at $\omega_p > 16.5$ eV could be obtained by various optical measurements. For example, by combining the absorption, reflectance or transmission measurements with Kramers-Kronig relationship, the frequency-dependent optical excitations could be verified.

## 4. Conclusions

We have performed first-principles calculations of the geometric, electronic properties and exitonic effect on optical spectrum of Lithium metasilicate. Under this state-of-the-art analysis, a concise physical/chemical picture and the crucial characteristics of the electrolyte in $Li^+$ related materials are achieved, through which the diversified essential properties of cathode/electrolyte/anode materials in ion-related materials in general could be fully understood. Lithium metasilicate exhibits special properties, a large Moire cell with extremely non-uniform physical/chemical environments, wide-indirect insulator gap, a significant spatial charge distribution generated by the orbital hybridizations, and atom-/orbital-projected density of states. Consequently, the distinct orbital hybridization related to the optical response could be identified. The outstanding optical properties involve the un-isotropic optical behaviors under three electronic dipole directions, the robust excitonic states with a largely reduced optical gap ($E_g^O = 6.95$ eV) much smaller than the band gap ($E_g^i = 8.4$ eV), strongly modified single particle interband transitions, the reflectance/high transparency of electromagnetic waves for $\omega < E_g^o$, a high absorption coefficient, twenty prominent single-particle excitation peaks, and many pronounced peaks in ELF at the plasmon resonance frequency $\omega_p$, $17\ eV - 19\ eV$. The main optical features of Li$_2$SiO$_3$ could be verified by the experimental optical spectroscopy measurements. The current theoretical framework could be developed for further studies of other anode/cathode/electrolyte materials for LIBs.

## Acknowledgments

This work is supported by the Hi-GEM Research Center and the Taiwan Ministry of Science and Technology under grant number MOST 108-2212-M-006-022-MY3, MOST 109-2811-M-006-505 and MOST 108-3017-F-006-003.

**Table 1**: Prominent absorption structures: Frequencies, colored indicators and identified orbital hybridizations.

| Excitation frequencies (eV) | | Colored arrows | Specific orbital hybridizations in Si-O bonds |
|---|---|---|---|
| With excitonic effect | Without excitonic effect | | |
| 7.8 | | Dash | |
| 8.4 | | Black with red border | |
| 9.0 | 9.0 | Dash red | Si(3s)-O(2p$_z$) |
| 9.5 | 9.5 | Dash green | Si(3s)-O (2p$_x$, 2p$_y$, 2p$_z$) |
| 10 | 10 | Dash dark yellow | Si(3s)-O (2p$_x$, 2p$_y$, 2p$_z$) |
| 10.4 | 10.4 | Dash brown | Si(3s)-O (2p$_x$, 2p$_y$, 2p$_z$) |
| 11 | 11 | Dash purple | Si(3s)-O (2p$_x$, 2p$_y$, 2p$_z$) |
| 11.4 | 11.4 | Dash black | Si(3s)-O (2p$_x$, 2p$_y$, 2p$_z$) |
| 12 | 12 | Dash orange | Si (3s, 3p$_x$, 3p$_y$, 3p$_z$)-O (2p$_x$, 2p$_y$, 2p$_z$) |
| 12.5 | 12.5 | Dash grey | Si(3s)-O (2p$_x$, 2p$_y$) |
| 12.8 | 12.8 | Cyan | Si (3s, 3p$_x$, 3p$_y$, 3p$_z$)-O (2p$_x$, 2p$_y$, 2p$_z$) |
| 13.4 | 13.4 | Blue | Si (3s, 3p$_x$, 3p$_y$, 3p$_z$)-O (2p$_x$, 2p$_y$, 2p$_z$) |
| 14.3 | 14.3 | Black | Si (3s, 3p$_x$, 3p$_y$, 3p$_z$)-O (2p$_x$, 2p$_y$, 2p$_z$) |
| 15.2 | 15.2 | Green | Si (3s, 3p$_x$, 3p$_y$, 3p$_z$)-O (2p$_x$, 2p$_y$, 2p$_z$) |
| 16 | 16 | Dark yellow | Si (3s, 3p$_x$, 3p$_y$, 3p$_z$)-O (2p$_x$, 2p$_y$, 2p$_z$) |
| 16.6 | 16.6 | Purple | Si (3s, 3p$_x$, 3p$_y$, 3p$_z$)-O (2p$_x$, 2p$_y$, 2p$_z$) |
| 17.5 | 17.5 | Pink | Si (3s, 3p$_x$, 3p$_y$, 3p$_z$)-O (2p$_x$, 2p$_y$, 2p$_z$) |
| 17.9 | 17.9 | Grey | Si (3s, 3p$_x$, 3p$_y$, 3p$_z$)-O (2p$_x$, 2p$_y$, 2p$_z$) |
| 19.2 | 19.2 | Red | Si (3s, 3p$_x$, 3p$_y$, 3p$_z$)-O (2p$_x$, 2p$_y$, 2p$_z$ |
| 20.2 | 20.2 | Orange | Si (3s, 3p$_x$, 3p$_y$, 3p$_z$)-O (2p$_x$, 2p$_y$, 2p$_z$ |
| 21 | 21 | Brown | Si (3s, 3p$_x$, 3p$_y$, 3p$_z$)- (2p$_x$, 2p$_y$, 2p$_z$) |
| 27.5 | 27.5 | Dark Blue | Si(3s)-O(2s) |

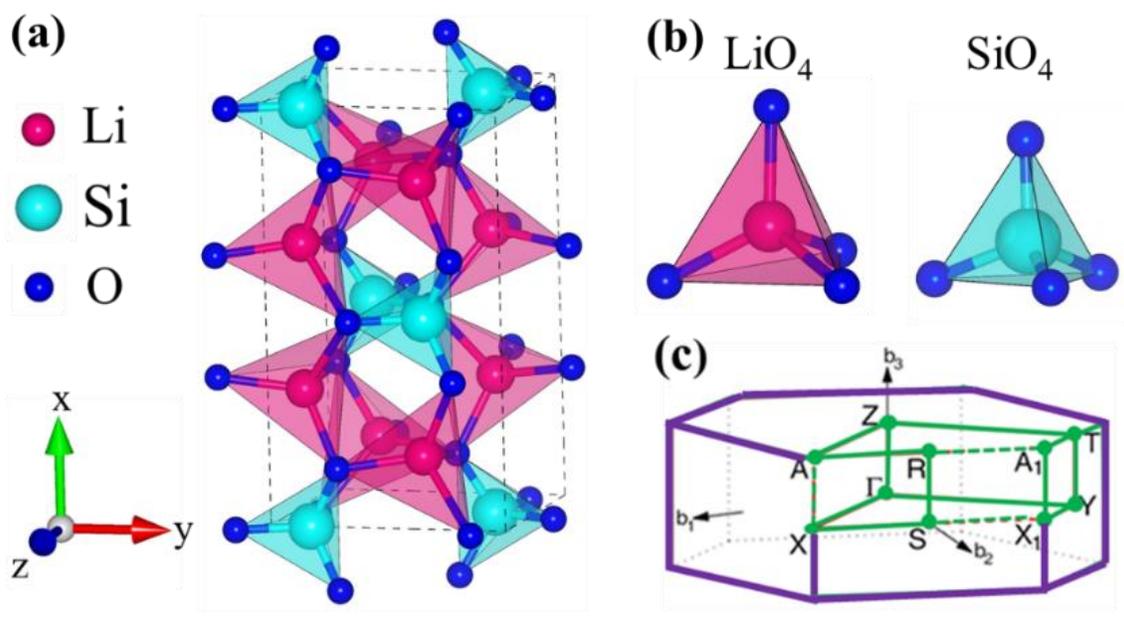

**Figure 1**. (a) The optimal geometric structure of the $Li_2SiO_3$ compound. The dashed black line indicates (b) oxygen atoms around each Li and Si atom, and (c) the first Brillouin zone along the high symmetry points.

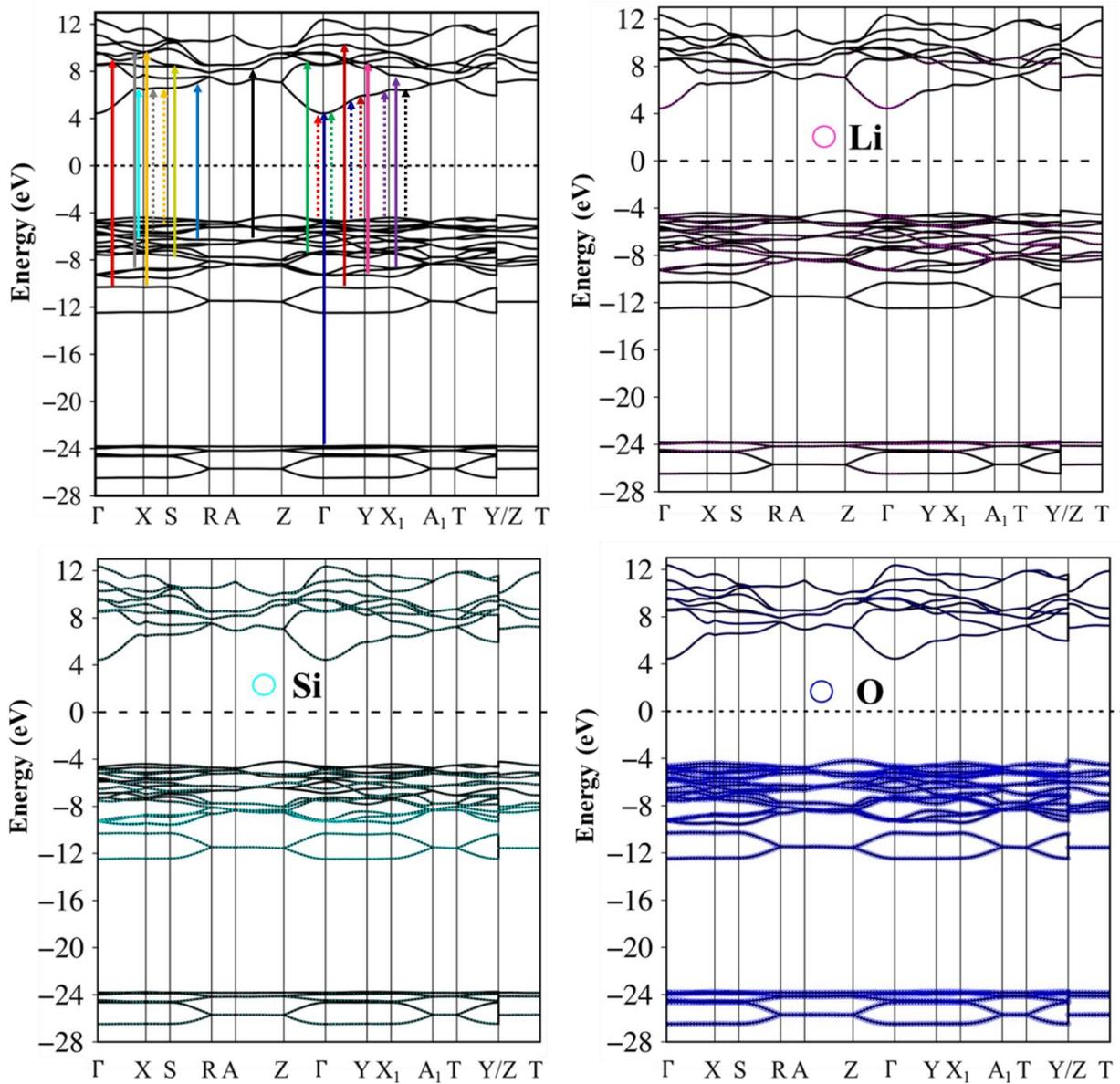

**Figure 2.** (a) Band structure along the high-symmetry points in the wave-vector space, with (b) Li-, (c) Si- and (d) O-atom dominances (pink, cyan and blue circles respectively)

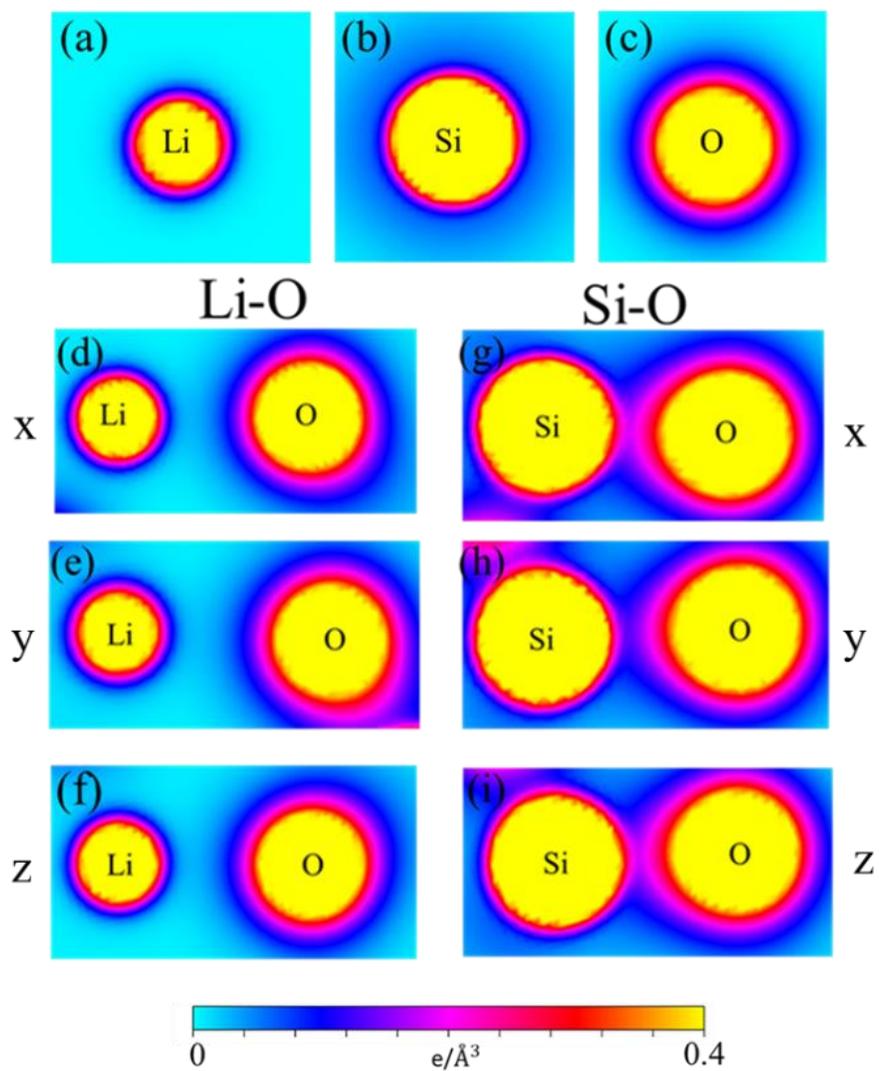

**Figure 3.** Comparison of the isolated Li/ Si/ O atoms respectively (a)/ (b)/ (c). The charge density distributions related to the significant orbital hybridizations in Li-O and Si-O bonds, respectively (d)/(e)/(f) and (g)/(h)/(i) along the x-/y-/z-directions for the shortest bonds.

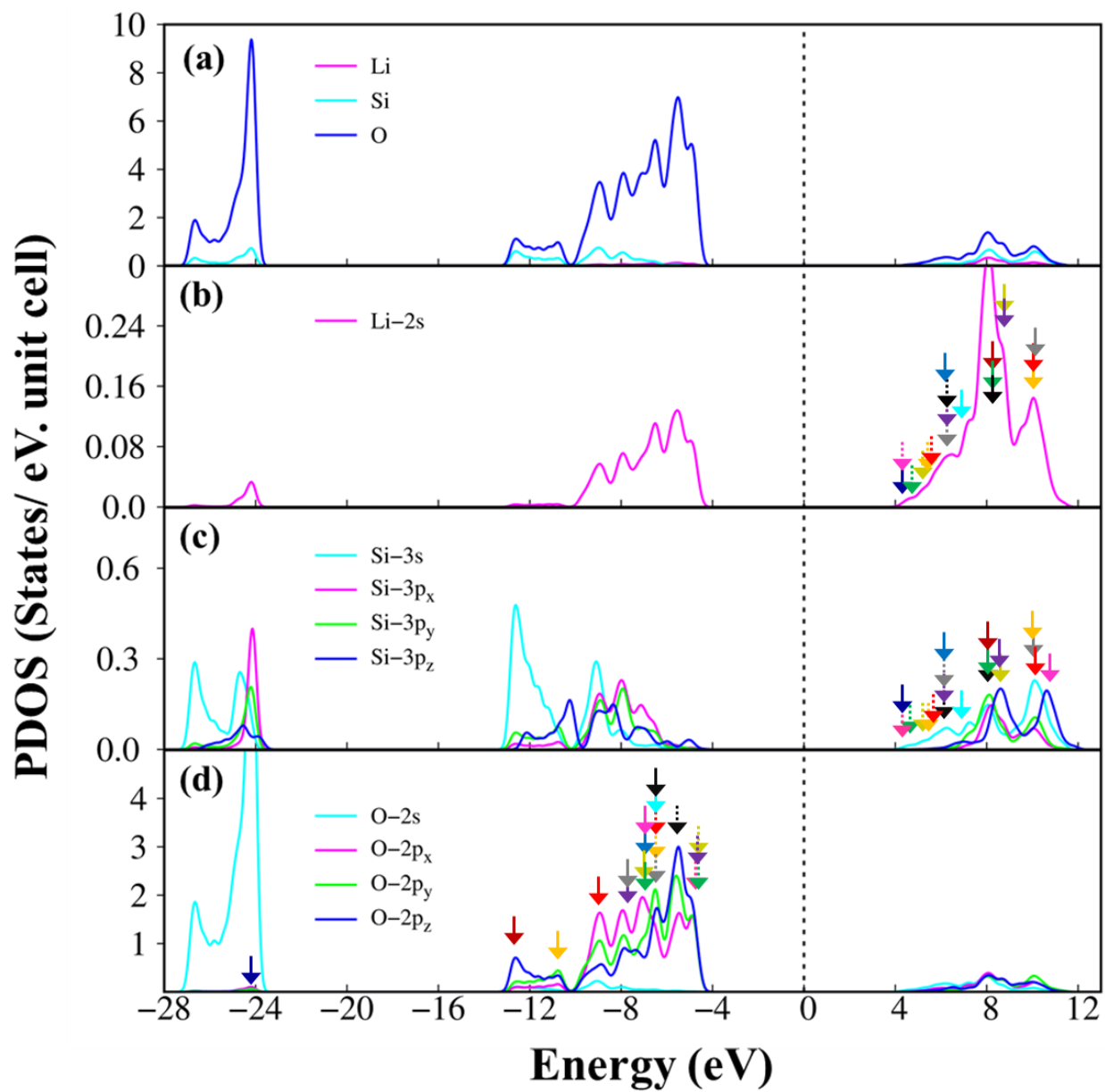

**Figure 4.** Density of states under the different components: (a) the total magnitude with Lithium, Silicon and Oxygen atom contributions, (b) Lithium and 2s-decomposed results, (c) Silicon and (3s, 3$p_x$, 3$p_y$, 3$p_z$)-projected calculations, and (d) Oxygen and (2s, 2$p_x$, 2$p_y$, 2$p_z$)-related ones.

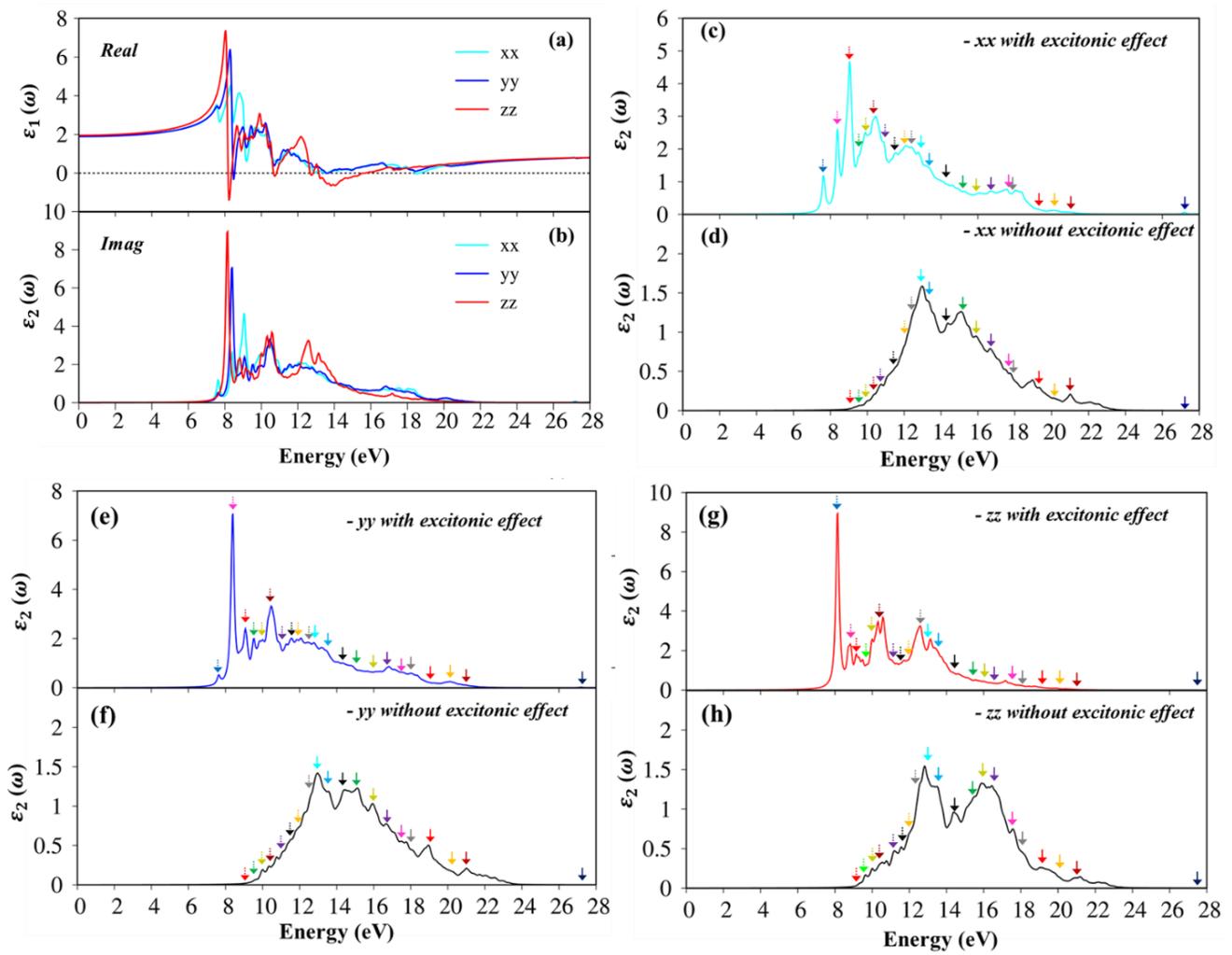

**Figure 5**. (a) The real-part and (b) imaginary-part dielectric functions with the excitonic effects under x-/y-/z- polarization directions (the cyan, blue and red curves). Comparison of the imaginary part of dielectric function (c), (d)/ (e), (f)/(g), (h) with and without excitonic effect under x-/y-/z- polarization directions, respectively.

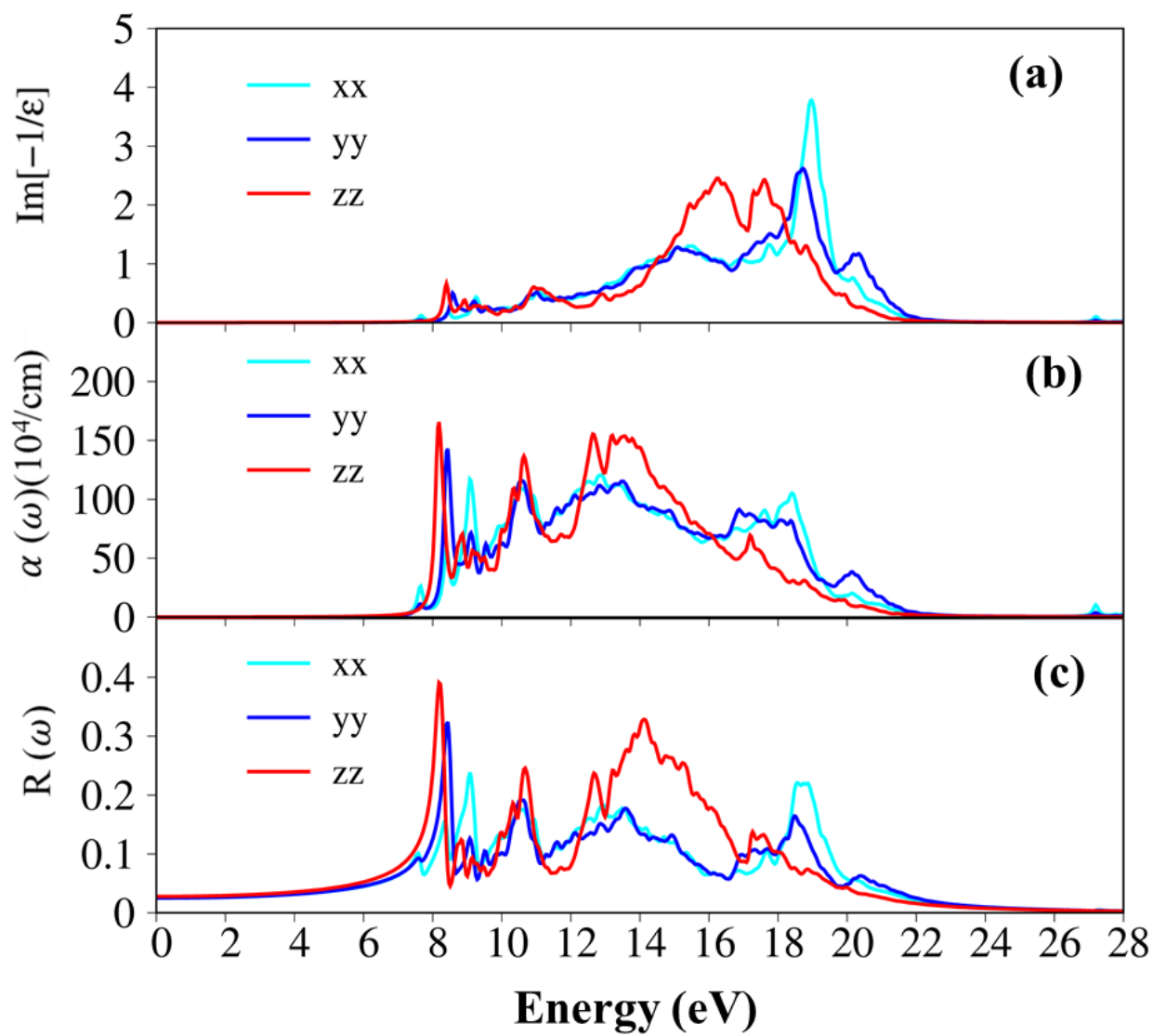

**Figure 6**. The various optical properties: (a) energy loss functions, (b) absorption coefficients and (c) reflectance spectra under the three electric-polarization directions (the cyan, blue and red curves)**.**